%% file: BKp-main.tex
\documentclass[12pt]{article}

\input definitions.tex
\renewcommand{\theequation}{\arabic{section}.\arabic{equation}}

\begin{document}
\title{
$~$
\vspace{-25mm}
\begin{flushright}
{\small LBNL-51261
\\
UCB-PTH-02/32
\\
\hfill hep-th/0207174
\\}
\end{flushright}
\vspace{5mm}
\textbf{On Continuous Moyal Product Structure in String Field Theory}
}
\author{
\textsf{D.M.~Belov}\footnote{On leave from Steklov Mathematical
Institute, Moscow, Russia.}
\vspace{2mm}
\\
Department of Physics
\\
Rutgers University
\\
136 Frelinghuysen Rd.
\\
Piscataway, NJ 08854, USA
\vspace{2mm}
\\
\texttt{belov@physics.rutgers.edu}
\vspace{0.5cm}
\\
\textsf{and}
\\
\textsf{A.~Konechny}
\vspace{3mm}
\\
Department of Physics,\\ University of California Berkeley \\
and
Theoretical Physics Group,\\
 Mail Stop 50A-5101\\
LBNL, Berkeley, CA 94720 USA
\vspace{2mm}
\\
\texttt{konechny@thsrv.lbl.gov}
\vspace{-10mm}
}
\date{~}

\maketitle
\thispagestyle{empty}

\begin{abstract}
We consider a diagonalization of Witten's star product for a ghost system
of arbitrary background charge and Grassmann parity. To this end we
use a bosonized formulation of such systems and a spectral analysis of
Neumann matrices. We further identify a continuous Moyal product structure
for a combined ghosts+matter system. The normalization of multiplication
kernel is discussed.
\end{abstract}

\newpage
%\dopage{\finkfile}

\tableofcontents

%%%%%%%%%%%%%%%%%%%%%%%%%%%%%%%%%%%%%%%%%%%%%%%%%%%%
%\clearpage
\section{Introduction}\label{sec:intro}
\dopage{\finkfile}
\setcounter{equation}{0}
%%%%%%%INTRO
A representation of Witten's star product in open string field theory  (SFT)
\cite{OSFT} as a family
of infinitely many Moyal products was first proposed by I.~Bars in \cite{bars1}.
The approach of that paper is essentially based on split formalism in SFT, it was further
elaborated in papers \cite{bars1}, \cite{bars2}.
A different approach to Moyal-type representation for   Witten's star product
 based on spectroscopy of Neumann matrices \cite{spectroscopy} was taken up in
 \cite{0202087}. The authors of that paper showed that in the matter sector the Witten's
 star can be represented as a continuous Moyal product specified  by a family of
 noncommutativity parameters
\begin{equation*}
\theta(\kappa)=2\tanh\left(\frac{\pi\kappa}{4}\right)
\end{equation*}
labelled by a continuous variable $0\le\kappa<\infty$.
An additional assumption in that paper was the restriction to zero momentum sector.
The whole construction was generalized to arbitrary momentum in \cite{dima1}.

One hopes that reformulation of string field theory that uses a continuous Moyal product
will make the whole structure more transparent and introduce new computational tools.
Given the fact that SFT axioms \cite{OSFT} mimic the conventional axioms of noncommutative
geometry \cite{Connes},
the theory written in some new basis may ultimately resemble a lot noncommutative
field theories (see \cite{Nikita}, \cite{KS}, \cite{ABGKM}, \cite{ohmori},
\cite{desmet} for a review).
Thus in noncommutative field theories solitonic solutions were explicitly
constructed \cite{GMS} using Moyal algebra projectors. It turns out the well
known sliver solution \cite{KP} in SFT being rewritten in the continuous Moyal
basis is represented by a functional
$$
\Xi[x_{\alpha}(\kappa)]= {\cal N}\exp\left[-\int_{0}^{\infty}d\kappa\,
\frac{x_{\alpha}(\kappa)x_{\alpha}(\kappa)}{\theta(\kappa)}  \right]
$$
that
has the form of a continuous family of noncommutative field theory projectors
(see \cite{sliver_cont} for a detailed discussion of this representation of the sliver).

To complete the program and rewrite the whole SFT (VSFT or Witten's cubic SFT)
in the continuous Moyal basis one needs to extend the formalism to
 include the ghost sector and a BRST operator.
In the present paper we extend the approach of \cite{0202087} to ghosts. A preliminary work
in that direction was done in \cite{dima1} where ghosts in the bosonic SFT were
treated in the fermionic $b$-$c$-representation. In \cite{Erler} a continuous
Moyal product  was constructed for  string fields including
ghosts (in $bc$ formalism) restricted to Siegel's gauge. In this paper we
work in the bosonized representation considering ghost systems of arbitrary background
charge and Grassmann parity.
 We find working in  bosonized representation technically simpler
than treating $bc$- or $\beta\gamma$-systems directly.
The continuous Moyal product for fermionic fields was considered in \cite{steklovka}.

The paper is organized as follows. In section 2 we set up some notations and
remind the reader the form of two and three-string vertices for ghosts in
the bosonized representation. In section 3 we diagonalize these vertices using
the spectroscopy of Neumannn matrices. In section 4 we represent the Witten's
star product as a functional integral operator specified by a certain kernel.
In section 5 we discuss regularization of determinants in the continuous
Moyal basis. As an instructive exercise we  check the
 cancellation of infinities for overlaps of surface states.
 In section 6 we apply the technique of section 5 to the normalization constant
 of the multiplication kernel. This section also contains further discussion
 of the precise correspondence between the Witten star product and the
 continuous Moyal one. We end with conclusions in section 7.
Appendices contains technical details of the calculations.

%%%%%%%%%%%%%%%%%%%%%%%%%%%%%%%%%%%%%%%%%%%%%%%%%%%%
%\clearpage
\section{Ghost $2$- and $3$-string vertices in the
bosonized formulation}
\label{sec:3string}
\setcounter{equation}{0}
\input{BKp-3string}

%%%%%%%%%%%%%%%%%%%%%%%%%%%%%%%%%%%%%%%%%%%%%%%%%%%%
%\clearpage
\section{Diagonalization}
\label{sec:diag}
\setcounter{equation}{0}
\input{BKp-diag}

%%%%%%%%%%%%%%%%%%%%%%%%%%%%%%%%%%%%%%%%%%%%%%%%%%%%
%\clearpage
\section{Multiplication Kernel}
\label{sec:mult}
\setcounter{equation}{0}
\input{BKp-mult}

\section{Overlaps of Surface States}
\label{sec:surstate}
\setcounter{equation}{0}
\input{BKp-surstate}

\section{Continuous Moyal  versus Witten's star product}
\label{sec:iso}
\setcounter{equation}{0}
\input{BKp-iso}

%\section{Surface states}
%\label{sec:surface}
%\setcounter{equation}{0}
%\input{BK-surface}

%\clearpage
\section{Conclusions}
\setcounter{equation}{0}
Here we summarize the main results of the paper.

 We diagonalized the $3$-string vertex for a general
bosonized Bose/Fermi ghost system (see \eqref{diagvert}).
Our results can be
easily generalized for the non-zero momentum matter sector
(see Appendix~\ref{appD}).
In the diagonal basis  the string star product was rewritten in a mixed coordinate/momentum
representation  and  expressions for the kernel
and its normalization constant (see \eqref{kernelQ}) were obtained. We showed
that this kernel defines an associative multiplication.
In particular our computations  reveal    that the
correct midpoint insertion operator is necessary for the associativity.

One of the side issues considered along the way  was a non-perturbative proof of the
statement that the $\bpz$ inner product of any
two surface states (with appropriate ghost insertions) is equal
to one in CFT with vanishing central charge.
We considered in detail the computation
of an overlap of the wedge state  $|0\rangle\star|0\rangle$ with itself.
It was shown that for the critical bosonic string
this inner product is finite. We also discussed calculation of finite parts in the
identity.

 The main issue considered in the paper is   an isomorphism
 between the Witten star product algebra and a
canonically normalized continuous Moyal algebra.
 An explicit expression for the constant relating
these to algebras was found \eqref{constant}. Modulo some potential subtleties
having to do with the finite part of spectral density this constant appears to
be divergent.

One of the unsolved problems among others discussed  in this paper is to
obtain an analytic expression for the finite
part of the spectral density \eqref{density}.

%%%%%%%%%%%%%%%%%%%%%%%%%%%%%%%%%%%%%%%%%%%%%%%%%%%%%%%%%%%
\section*{Acknowledgments}
We would like to thank H.~Liu and B.~Zwiebach for useful discussions.
D.~Belov would like to acknowledge the hospitality of the
Lawrence Berkeley National Laboratory.
The work of D.B. was supported by DOE grant DE-FG02-96ER40959
and in part by RFBR grant 02-01-00695.
The work of A.~Konechny was supported by the Director, Office of Energy
Research, Office of High Energy and Nuclear Physics, Division of High
Energy Physics of the U.S. Department of Energy under Contract
DE-AC03-76SF00098 and in part by the National Science
Foundation grant PHY-95-14797.

%%%%%%%%%%%%%%%%%%%%%%%%%%%%%%%%%%%%%%%%%%%%%%%%%%%%%%%%%%%%%%%
% APPENDIX
%%%%%%%%%%%%%%%%%%%%%%%%%%%%%%%%%%%%%%%%%%%%%%%%%%%%%%%%%%%%%%%
\clearpage
\appendix
\section*{Appendix}
\addcontentsline{toc}{section}{Appendix}
\renewcommand {\theequation}{\thesection.\arabic{equation}}

\section{Relations involving vector $J_n$}
\label{appA}
\setcounter{equation}{0}
\input{BKp-appA}

\section{Diagonalization of $3$-string vertex}
\label{appB}
\setcounter{equation}{0}
\input{BKp-appB}

\section{Calculations of the kernel}
\label{appC}
\setcounter{equation}{0}
\input{BKp-appC}

\section{Non-zero momentum matter $3$-string vertex}
\label{appD}
\setcounter{equation}{0}
\input{BKp-appD}

%%%%%%%%%%%%%%%%%%%%%%%%%%%%%%%%%%%%%%%%%%%%%%%%%%%%%%%%%%
\newpage
%\dopage{\finkfile}

\end{document}

%% file: definitions.tex
\usepackage{amssymb,amsmath,amsthm}
\usepackage{array,longtable}
\usepackage[dvips]{graphicx}
\usepackage[dvips]{color}

\usepackage[mathscr]{eucal}

\newcommand{\finkfile}{}

\newcolumntype{V}{>{$}m{4cm}<{$}}
\newcolumntype{C}{>{$}c<{$}}
\newcolumntype{L}{>{$}l<{$}}
\newcolumntype{R}{>{$}r<{$}}

\newcommand{\af}{a}
\newcommand{\of}{o}
\newcommand{\ef}{e}

\newcommand{\Ds}{\mathscr{D}}
\newcommand{\Fs}{\mathscr{F}}

\newcommand{\Cc}{\mathcal{C}}
\newcommand{\Rh}{\mathbb R}
\newcommand{\Zh}{\mathbb Z}

\newcommand{\Ks}{\mathscr{K}}
\newcommand{\Nc}{\mathcal{N}}

\newcommand{\Hc}{\mathcal{H}}

\newcommand{\pd}{\partial}

\newcommand{\oz}{\overline{z}}

\newcommand{\Tr}{\mathop{\mathrm{Tr}}\nolimits}

\newcommand{\bpz}{\mathop{\mathrm{bpz}}\nolimits}

\allowdisplaybreaks[3]

%% file: BKp-3string.tex
%\input definitions.tex
%\begin{document}

\dopage{\finkfile}

\subsection{Bosonized  formulation}
In bosonized formulation a ghost $bc$- or $\beta\gamma$-
system  is represented by a single bosonic field $\phi(z,\bar z)$
with a  mode expansion \cite{fms,Friedan}
\begin{subequations}
\begin{equation}
\epsilon\phi(z,\oz)=2i\phi_0+j_0\log z\oz -\sum_{n\ne 0}\frac{j_n}{n}\left(
\frac{1}{z^n}+\frac{1}{\oz^n}\right)
\end{equation}
where $j_0$ is the ghost number operator taking integral values $q\in {\mathbb Z}$,
parameter $\epsilon$ takes values $+1$ for a fermionic $b$-$c$ system and  $-1$ for
a bosonic $\beta$-$\gamma$ ghost system. The parameter $\epsilon$ also enters
the  commutation relations
\begin{equation}
[\phi_0,j_0]=i\epsilon
\quad\text{and}\quad
[j_n,j_m]=\epsilon n\delta_{n+m,0}\, .
\end{equation}
A conformal stress-energy tensor for a field $\phi$ with a background charge $Q$
is given by the expression
\begin{equation}
T_{\phi}=\frac{\epsilon}{2}\pd\phi\pd\phi-\frac{Q}{2}\pd^2\phi.
\end{equation}
\label{phi}
\end{subequations}
The last term in the stress energy tensor
is due to the anomaly in the conservation of the ghost current $j=\epsilon\pd\phi$.
The number  $Q$ is a background charge, which is equal
to $-3$ for conformal  $bc$ ghost system of the bosonic string.

 Denote by $|q\rangle$ a state that is
annihilated by $j_n$, $n>0$ and is of ghost number $q$:
\begin{equation}
j_n|q\rangle=0\quad\text{and}\quad j_0|q\rangle = q|q\rangle.
\end{equation}
Because of the anomaly in the conservation of the ghost
charge the inner product is nonzero only for the following states
\begin{equation}
\langle -q-Q|q'\rangle=\delta_{qq'}.
\end{equation}

We further introduce creation and annihilation operators according to
\begin{subequations}
\begin{equation}
a_{n}^{\dagger}= \sqrt{n} j_{-n}\quad
\text{and}\quad a_{n}=\sqrt{n}j_{n}\, ,
\enspace n>0 \, .
\end{equation}
Notice that the commutation relations for these  oscillators
 include $\epsilon$
\begin{equation}
[\af_n,\,\af^{\dag}_m]=\epsilon \delta_{nm}.
\end{equation}
\end{subequations}

\subsection{Overlap vertices}
 The two operations: multiplication and inner product,
 needed to define string field theory action,
can be written in terms of $3$- and $2$-string vertices.
Here we will present only the vertices for the bosonized
ghosts. The ones for the matter part can be found,
for example in \cite{GJ1,peskin1,0102112}.

The 3-string vertex for a bosonized field (\ref{phi}) can be written as
\cite{peskin2,Jevicki}
\begin{subequations}
\begin{multline}
|V_3^{(\phi)}\rangle_{123}=\sum_{q_r\in\Zh}\delta_{q_1+q_2+q_3+Q,0}
\exp\left[
-\frac{\epsilon}{4}V_{00}^{\prime}\Bigl(\sum_r q_r^2-Q^2\Bigr)
\right.
+\frac{\epsilon}{\sqrt{2}}
\sum_{r,s}\sum_{n=1}^{\infty}q_r V_{0n}^{\prime rs}\af_n^{(s)\dag}
\\
-\frac{\epsilon}{2}\sum_{r,s}\sum_{n,m=1}^{\infty}
\af_n^{(r)\dag}V_{nm}^{\prime rs}\af_m^{(s)\dag}
\left.
 -\epsilon \frac{Q\sqrt{2}}{6}\sum_{r=1}^{3}\sum_{n=1}^{\infty}J_{n}\af_n^{(r)\dag}\right]
\bigotimes_{r=1}^3
| -q_r-Q\rangle,
\label{V3}
\end{multline}
where
\begin{align}
V_{nm}^{\prime rs}&=\frac13\bigl(
C'+\alpha^{s-r}U^{\prime}+\alpha^{r-s}\bar{U}^{\prime}
\bigr)_{nm}
\\
V_{0m}^{\prime rs}&=\frac13\bigl(
\alpha^{s-r}W^{\prime}_m+\alpha^{r-s}\bar{W}^{\prime}_m
\bigr)
\\
V_{00}^{\prime}&=-2\log\gamma,\qquad \gamma=\frac{4}{3\sqrt{3}}
\end{align}
are the standard $3$-string vertex matrices
\cite{GJ1,0102112}, $C_{mn}^{\prime}$ is a twist matrix
$(-1)^n\delta_{nm}$ and vector  $J_n$ is given by
\begin{equation}
J_{2n}=\frac{(-1)^n}{\sqrt{n}}\quad\text{and}\quad J_{2n+1}=0\,.
\label{Jn}
\end{equation}
\end{subequations}
The indices $r$ and $s$ in (\ref{V3}) run from $1$ to $3$ and label the
tensorial components in the 3-string Hilbert space. Essentially (\ref{V3})
differs from the matter vertex by the presence of the background charge in the
Kronecker symbol and the term linear in $Q$ in the exponent which arises
from the insertion of the operator \cite{GJ1}
\begin{equation}
\exp\left[-\frac{\epsilon Q}{12}\sum_{r=1}^{3}\phi^{(r)}\Bigl(\frac{\pi}{2}\Bigr)\right]
\label{inser}
\end{equation}
into the vertex \eqref{V3} with $Q=0$\footnote{
The mid-point insertion for $3$-string vertex in  notations of \cite{GJ1}
is equal to
$$
\exp\left[-\frac{iQ}{2}\frac{1}{3}\sum_{r=1}^{3}\phi^{(r)}_{\text{GJ}}
\bigl(\frac{\pi}{2}\bigr)\right].
$$
To obtain expression \eqref{inser} we
take into account that $\phi_{\text{GJ}}=-\frac{i}{2}\phi_{\text{here}}$.
}.

As a side remark we would like to note that
the exponent here is not normal ordered.
Formally, if one acts by
the operator \eqref{inser} on the vertex \eqref{V3} with $Q=0$,
one obtains an extra numeric factor
\begin{equation*}
\exp\left[-\frac{\epsilon Q^2}{24}\sum_{n=1}^{\infty}J_n^2\right]\, .
\end{equation*}
This factor is important for the overall normalization of the vertex,
our normalization coincides with the one in \cite{peskin2}.
Notice also that the quadratic term in the vertex exponential
is exactly the same as that of
the matter part upon the substitution $\epsilon\mapsto g_{\mu\nu}$.

The 2-string vertex reads \cite{peskin2}
\begin{equation}\label{V2}
|V_{2}^{(\phi)}\rangle_{12} = \sum_{q\in {\mathbb Z}}
\exp\left[-\epsilon\sum_{n=1}^{\infty}
\af_{n}^{(1)\dagger}(-1)^{n}\af_{n}^{(2)\dagger}
\right]|q\rangle_{1}\otimes |-q-Q\rangle_{2} \, .
\end{equation}
It realizes the $\bpz$ conjugation on the modes of field $\phi$.

%\end{document}

%% file: BKp-diag.tex
%\input definitions.tex
%\begin{document}

\dopage{\finkfile}

\subsection{Notations}
The diagonalization of the matter vertex (at zero momentum) found in
\cite{0202087} is based on the spectral analysis of Neumann matrices done in
\cite{spectroscopy}. It was shown in \cite{spectroscopy} that the matrices
$(C'V')_{mn}^{ rs}$ have a joint set of eigenvectors $v_{n}^{(\kappa)}$ labeled
by a continuous parameter $-\infty<\kappa<\infty$. Namely we have
\begin{equation*}
\sum_{n}(C'V^{\prime rs})_{mn}v_{n}^{(\kappa)}=\mu^{rs}(\kappa)v_{m}^{(\kappa)}
\end{equation*}
where
\begin{equation}
\mu^{rs}(\kappa)=\frac{1}{1+2\cosh\frac{\pi\kappa}{2}}
\Bigl[1-2\delta_{r,s}+e^{\frac{\pi\kappa}{2}}\delta_{r+1,s}
+e^{-\frac{\pi\kappa}{2}}\delta_{r,s+1}\Bigr].
\end{equation}

The vectors $v^{(\kappa)}$ are chosen to satisfy the following orthogonality and
completeness relations \cite{okuyama1}
\begin{equation}
(v^{(\kappa)},v^{(\kappa')})=\Nc(\kappa)\delta(\kappa-\kappa')
\quad\text{and}\quad
\int_{-\infty}^{+\infty} \frac{d\kappa}{\Nc(\kappa)}\,
v_{m}^{(\kappa)}v_{n}^{(\kappa)}=\delta_{mn}
\end{equation}
where
\begin{equation}
\Nc(\kappa)=\frac{2}{\kappa}\sinh\left( \frac{\pi\kappa}{2}\right) \, .
\end{equation}

We use  $v^{(\kappa)}$ to  introduce
a new oscillator basis
\begin{equation}
\af_{\kappa}^{\dag}=\frac{1}{\sqrt{\Nc(\kappa)}}\sum_{n=1}^{\infty}\af_{n}^{\dag}
v^{(\kappa)}_n\,
\end{equation}
and the same for $\af_{\kappa}$.
The action of the operator $C'$ and commutation relations are of the form
\begin{equation}
C'\af_{\kappa}^{\dag}=-\af_{-\kappa}^{\dag}\quad\text{and}\quad
[\af_{\kappa},\,\af_{\kappa'}^{\dag}]=\epsilon\delta(\kappa-\kappa').
\end{equation}
We next perform one more change of the oscillator basis which
diagonalizes  the  operator $C'$
\begin{subequations}
\label{ae}
\begin{alignat}{3}
\ef_{\kappa}^{\dag}&=\frac{\af_{\kappa}^{\dag}+C'\af_{\kappa}^{\dag}}{\sqrt{2}},
&\qquad
\of_{\kappa}^{\dag}&=\frac{\af_{\kappa}^{\dag}-C'\af_{\kappa}^{\dag}}{i\sqrt{2}}
&\quad \kappa &\geqslant 0; \label{o_e}
\\
\af_{\kappa}^{\dag}&=\frac{\ef_{\kappa}^{\dag}+i\of_{\kappa}^{\dag}}{\sqrt{2}},
&
C'\af_{\kappa}^{\dag}&=\frac{\ef_{\kappa}^{\dag}-i\of_{\kappa}^{\dag}}{\sqrt{2}}
& \kappa &\geqslant 0.
\end{alignat}
In addition one has
\begin{equation}
\of_{-\kappa}^{\dag}=\of_{\kappa}^{\dag}\quad\text{and}
\quad \ef_{-\kappa}^{\dag}=-\ef_{\kappa}^{\dag}\, .
\end{equation}
\end{subequations}
Notice that for $\kappa=0$ there is only one non-vanishing mode $o_{\kappa=0}^{\dag}$.
The $\bpz$ conjugation specified by (\ref{V2}) acts on these oscillators as
\begin{equation}\label{bpz}
\bpz \ef_{\kappa}=- \ef_{\kappa}^{\dag}\quad\text{and}
\quad \bpz \of_{\kappa}=- \of_{\kappa}^{\dag}
\end{equation}
and the commutation relations are
\begin{equation}
[\ef_{\kappa},\,\ef_{\kappa'}^{\dag}]=\epsilon\delta(\kappa-\kappa')
\quad\mbox{and}\quad
[\of_{\kappa},\,\of_{\kappa'}^{\dag}]=\epsilon\delta(\kappa-\kappa') \, .
\end{equation}

\subsection{$3$-string and $2$-string vertices}
A straightforward computation yields the following form of the
$3$-string vertex in the new basis (see Appendix~\ref{appB} for details)
\begin{multline}
|V_{3}^{(\phi)}\rangle_{123}=\sum_{q_r}\delta_{q_1+q_2+q_3+Q,0}
\exp\left[\int_{0}^{\infty}d\kappa\left(
-\frac{\epsilon}{4}J_{\kappa}^2(1+3\mu)\Bigl[
\sum_r q^{2}_r-Q^2
\Bigr]\right. \right. \\
\left.\left.
%-\frac{\epsilon Q^2}{12}J_\kappa^2
-\epsilon\sum_r\Phi^{(r)}_{\alpha}(\kappa)\af_{\kappa,\alpha}^{(r)\dag}
-\frac{\epsilon}{2}\af_{\kappa,\alpha}^{(r)\dag}V^{rs}_{\kappa,\alpha\beta}
\af_{\kappa,\beta}^{(s)\dag}
\right)
\right] \bigotimes_{r=1}^{3}|-q_r-Q\rangle,
\label{diagvert}
\end{multline}
where the quadratic part in the exponent is given by an operator
\begin{subequations}
\begin{eqnarray}
&& V^{rs}_{\kappa,\alpha\beta}=\mu \delta_{\alpha\beta}\otimes \delta^{rs}
+\mu_s \delta_{\alpha\beta}\otimes\varepsilon^{rs}
+i\mu_a\epsilon_{\alpha\beta}\otimes\chi^{rs}\, ,\\
&& \mu_s=\frac12(\mu^{12}+\mu^{21})\, , \qquad \mu_a=\frac12(\mu^{12}-\mu^{21})
, \qquad \mu\equiv \mu^{11}
\end{eqnarray}
and we introduced the following  matrices
\begin{equation}\label{matrices}
\varepsilon^{rs}=
\left( \begin{array}{ccc}
0 & 1 & 1\\
1 & 0 & 1\\
1 & 1 & 0
\end{array}\right)\, ,
\qquad
\chi^{rs}= \left(
\begin{array}{ccc}
0 & 1 & -1\\
-1 & 0 & 1\\
1 & -1 & 0
\end{array}\right)\, ,
\qquad
\epsilon_{\alpha\beta}=\left(
\begin{array}{cc}
0 & 1 \\
-1 & 0
\end{array}\right) \, .
\end{equation}
The matrices $1,\varepsilon$ and $\chi$ form a closed algebra
\begin{equation}
\chi^2=\varepsilon-2,\qquad \varepsilon^2=\varepsilon+2,\qquad
\chi\varepsilon=\varepsilon\chi=-\chi.
\end{equation}
The indices $\alpha$, $\beta$ in the above expressions take values ``$e$''
and ``$o$'' for
even and odd modes (\ref{o_e}) respectively. Thus for instance
$\af^{(r)\dagger}_{\kappa,o}\equiv \of_{\kappa}^{(r)\dagger}$ and
$\af^{(r)\dagger}_{\kappa,e}\equiv \ef_{\kappa}^{(r)\dagger}$.
Since the transformation
$\af_n\to \af_{\kappa,\alpha}$ is
an orthogonal one, the vacuum state $|-q-Q\rangle$ in (\ref{diagvert})
stands for the tensor product of
$\af_{\kappa,\alpha}^{(r)}$-oscillators vacua and the state with the
ghost number $-q-Q$ from  the
Hilbert space corresponding to the operators $\phi_0,j_0$.
The constant terms in (\ref{diagvert}) and the terms linear in the creation operators
are expressed via
\begin{align} \label{Phi}
\Phi_{e}^{(r)}(\kappa)&=\frac{1}{2}J_{\kappa}(1+3\mu)\Bigl(
q^{(r)}+\frac{Q}{3}
\Bigr)+\frac{1}{3}J_{\kappa}Q \\
\Phi_{o}^{(r)}(\kappa)&=i J_{\kappa}\mu_a q^{(s)}\chi^{sr}
\equiv\frac{2i}{\theta(\kappa)}\,\Phi_e^{(s)}\chi^{sr}
 \, ,
 \label{Phioe}\\
J_{\kappa}&= \frac{1}{\sqrt{\Nc(\kappa)}}\,(v^{(\kappa)},J)=
\frac{\sqrt{2}}{\kappa\sqrt{\Nc(\kappa)}} \, .
\label{Jk}
\end{align}
\end{subequations}

%\end{document}

%% file: BKp-mult.tex
%\input definitions.tex
%\begin{document}

\dopage{\finkfile}

\subsection{Coordinate basis}
We now want to go to the coordinate representation corresponding to modes
$\ef_{\kappa}$, $\of_{\kappa}$. The coordinate eigenstates
are
\begin{subequations}
\label{basisX}
\begin{equation}
\langle X,q|=
\langle q|\exp\left[\int_{0}^{\infty}d\kappa \sum_{\alpha}\left(
-\frac{\epsilon}{2}\,x_{\alpha}(\kappa)x_{\alpha}(\kappa) +i\epsilon\sqrt{2}\,
\af_{\kappa,\alpha}x_{\alpha}(\kappa) +\frac{\epsilon}{2}\,
\af_{\kappa,\alpha}\af_{\kappa,\alpha}
\right)
\right],
\label{coordL}
\end{equation}
where $\langle q|$ stand for the tensor product of
$a_{\kappa,\alpha}^{(r)}$-oscillators bra-vector vacua with ghost number $q$.
The bpz-conjugated state is then given by the expression
\begin{equation}
|X,q\rangle=\exp\left[ \int_{0}^{\infty}d\kappa \sum_{\alpha}\left(
 -\frac{\epsilon}{2}x_{\alpha}(\kappa)x_{\alpha}(\kappa) - i\epsilon\sqrt{2}
 \af_{\kappa,\alpha}^{\dagger}x_{\alpha}(\kappa)
 +\frac{\epsilon}{2}\af_{\kappa,\alpha}^{\dagger}\af_{\kappa,\alpha}^{\dagger}
 \right)\right] |q\rangle
 \, .
\label{coord_kappa}
\end{equation}
\end{subequations}

 One can rewrite the coordinate state \eqref{coord_kappa}
in terms of the original  discrete oscillator basis $\af_n^{\dag}$.
To this end we introduce a vector
\begin{subequations}
\begin{equation}
X_n=\int_{-\infty}^{\infty}\frac{d\kappa}{\Nc(\kappa)}\,v_{n}^{(\kappa)}X_{\kappa},
\end{equation}
where
\begin{equation}
X_{\kappa}=\frac{x_e(\kappa)+ix_o(\kappa)}{\sqrt{2}}
\quad\text{and}\quad
-X_{-\kappa}=\frac{x_e(\kappa)-ix_o(\kappa)}{\sqrt{2}}.
\end{equation}
\end{subequations}
To understand the meaning of the vector $X_n$ it is fruitful
to compare it with notations of \cite{0202087}. Using the
fact that $x_{e}(\kappa)$ here is equal to $x(\kappa)$ in \cite{0202087}
and the same for $x_{o}(\kappa)$ here and $y(\kappa)$ there,
one can obtain the following relations
\begin{equation}
X_{2n}=\sqrt{2n}\, x_{2n}\quad\text{and}\quad
X_{2n-1}=-\frac{i}{\sqrt{2n-1}}\,p_{2n-1},
\end{equation}
where $x_n$ and $p_n$ are Fourier modes for
the field $\phi(\sigma)$ and its momentum $P(\sigma)$
\begin{equation*}
\epsilon\phi(\sigma)=2i\phi_0+\sqrt{2}\sum_{n=1}^{\infty}x_n\cos n\sigma
\quad\text{and}\quad
\epsilon\pi P(\sigma)=j_0+\sqrt{2}\sum_{n=1}^{\infty}p_n\cos n\sigma.
\end{equation*}

One can check then that the state \eqref{coord_kappa} takes the following
form in terms of oscillators $\af_n^{\dag}$ and the coordinates
$x_{2n}$ and $p_{2n-1}$
\begin{multline}
|X,q\rangle=\exp\left[\sum_{n=1}^{\infty}\left(
 -\frac{\epsilon}{2}(2n)x_{2n}^2 - i\epsilon\sqrt{2}\,
 \af_{2n}^{\dagger}\sqrt{2n}\,x_{2n}
 +\frac{\epsilon}{2}\af_{2n}^{\dagger}\af_{2n}^{\dagger}
 \right)\right]
\\
\times\exp\left[\sum_{n=1}^{\infty}\left(
 -\frac{\epsilon}{2}\frac{1}{2n-1}p_{2n-1}^2
 +\frac{\epsilon\sqrt{2}}{\sqrt{2n-1}}\,
 \af_{2n-1}^{\dagger}p_{2n-1}
 -\frac{\epsilon}{2}\af_{2n-1}^{\dagger}\af_{2n-1}^{\dagger}
 \right)\right]
  |q\rangle
 \, .
\label{coord_n}
\end{multline}
The calculations of the inner product of two such states yields
\begin{equation*}
\langle X,-q-Q|X',q\rangle=
\prod_{n=1}^{\infty}
\left[\frac{\pi}{2n\epsilon}\right]^{\frac12}\delta(x_{2n}-x_{2n}')
\times\prod_{n=1}^{\infty}
\left[\frac{\pi(2n-1)}{\epsilon}\right]^{\frac12}
\delta(p_{2n-1}-p_{2n-1}')\, .
\end{equation*}
Using this expression we can define the measure $\Ds X$ as
\begin{equation}
\Ds X=\prod_{n=1}^{\infty}\frac{dx_{2n}dp_{2n-1}}{\epsilon\pi}
\left[\frac{2n-1}{2n}\right]^{-\frac12}
\label{measure}
\end{equation}
 This measure coincides with the one used in \cite{0202087}.

With these notations we have the following representation for a unit operator in
the ghost number $q$ component of the Hilbert space ${\cal H}_{q}$
\begin{equation}
\textrm{Id}_{q}=\int \mathscr{D} X\, |X,q\rangle\otimes \langle X,-q-Q|\, .
\label{id}
\end{equation}

\subsection{String product in the coordinate basis}
Using  representation \eqref{id} for the identity operator
we can rewrite the string multiplication in the coordinate basis.
In this representation the product is realized as a product of wave functionals
$\Psi_q[X]$,
which are related to the states from $\Hc_q$ via
\begin{equation}
\Psi_q[X]=\langle \Psi_q|X,-q-Q\rangle.
\end{equation}
Then the multiplication of two functionals $\Psi_{q_1}$ and
$\Psi_{q_2}$ with  ghost numbers $q_1$ and $q_2$ respectively
gives a functional $\Psi_{q_1}\star\Psi_{q_2}$
of ghost number $q_1+q_2$  given by
\begin{equation}
(\Psi_{q_1}\star\Psi_{q_2})[X^{(3)}]
=\int \mathscr{D} X^{(1)}\mathscr{D} X^{(2)}\,
K_{q_1,q_2}(X^{(1)},X^{(2)},X^{(3)})\Psi_{q_1}[X^{(1)}]\Psi_{q_2}[X^{(2)}].
\end{equation}
Here the multiplication kernel $K_{q_1,q_2}$ is given by  the appropriate overlaps of the
$3$-string vertex with the coordinate eigenstates (see Appendix~\ref{appC} for
details)
\begin{subequations}
\label{kernelQ}
\begin{multline}
K_{\{q_1,q_2\}}(X^{(r)})\equiv\Bigl(\langle X^{(1)},q_1|\otimes
\langle X^{(2)},q_2|\otimes\langle X^{(3)},-Q-q_1-q_2|\Bigr)|V_3\rangle
\\
=
\Ks_Q
\exp\left[
2i\epsilon\int_0^{\infty}\frac{d\kappa}{\theta(\kappa)}\,
x^{(r)}_{e}(\kappa)\chi^{rs} x^{(s)}_{o}(\kappa)
\right]\times
\\
\exp\left[-i\epsilon\int_0^{\infty}d\kappa\,
\sqrt{2}J_{\kappa}\left\{ x_e^{(1)}(\kappa)
\Bigl(q_1+\frac{Q}{2}\Bigr) + x_e^{(2)}(\kappa)\Bigl(
q_2+\frac{Q}{2}\Bigr) - x_e^{(3)}(\kappa)\Bigl(
\bar{q}_3+\frac{Q}{2}\Bigr)\right\}
\right]
\label{kernel_q}
\end{multline}
where $\bar{q}_3=q_1+q_2$, the non-commutativity parameter $\theta(\kappa)$ is
\begin{equation}\label{theta}
\theta(\kappa)=2\tanh\left(\frac{\pi \kappa}{4}  \right)
\end{equation}
and $\Ks_Q$ is a normalization constant that can be formally written as
\begin{equation}
\Ks_Q=\frac{1}{\det 2(1+3\mu(\kappa))}\,\exp\left[
\frac{\epsilon Q^2}{24}\sum_{n=1}^{\infty}J_n^2-\frac{\epsilon Q^2}{3}\log\gamma
\right]\, .
\label{Kdet}
\end{equation}
\end{subequations}
We will discuss this constant in greater detail in the forthcoming sections.

Let us notice here that the multiplication kernel for the
bosonized ghosts differs from the one for the matter part (with $D=1$)
only by the presence of symbol $Q$ in the exponent. Essentially
if one considers the kernel for the matter sector with \textit{non-zero}
momentum one will obtain formula \eqref{kernel_q} with $Q=0$.
For future use we remind the expression for zero momentum kernel
in the matter sector
\begin{equation}
K^{\text{matter}}(X^{(r)})=
\Ks^{D}_{Q=0}\exp\left[ \int_{0}^{\infty}d\kappa\,\frac{2i}{\theta(\kappa)}\sum_{\mu=0}^{D-1}
x_{\mu,e}^{(r)}\chi^{rs}x_{\mu,o}^{(s)}\right].
\label{M}
\end{equation}

It is clear from the $+$, $+$, $-$ structure of the additional exponential
term in
the ghost kernel (\ref{kernel_q}) that it still defines an associative multiplication
in the ghost sector and moreover this additional factor
can be removed by a wave-function redefinition:
\begin{equation}\label{redef}
\tilde \Psi_{q}[X]=\exp\left[ -i\sqrt{2}\epsilon\Bigl(q+\frac{Q}{2}
\Bigr)\int_{0}^{\infty}d\kappa\,
 J_{\kappa} x_{e}(\kappa)\right]\Psi_{q}[X]\, .
\end{equation}
 The wave functions $\tilde \Psi_{q}(X)$ are multiplied with the help of
  the kernel \eqref{M} with $D=27$,
  where $\mu=26$ stands for the bosonized ghosts fields,
  and  the normalization constant should be $\Ks_{Q=-3}\Ks_{Q=0}^{26}$.

The rest of the paper is essentially devoted to the study of the normalization constant
(\ref{Kdet}). The total constant $\Ks_{Q=-3}\Ks_{Q=0}^{26}$
 contains two formally divergent factors. One comes from a determinant
of a certain operator and the other one, proportional to the background charge $Q^{2}$
originates from the ghost anomaly insertion in the vertex.
It was suggested in \cite{0202087} that in the critical bosonic string theory, i.e.
when $D=26$, $Q=-3$, $\epsilon=1$ and the combined central charge vanishes, the
two terms may cancel each other.

Indeed cancellations of this form occur in SFT \cite{peskin2}, \cite{sen_as}.
In the next section we consider in detail
 one example of such cancellation that has to do with overlaps of surface states.
 We will discuss a general method for regularization of infinite determinants of operators
 written in
 the continuous basis $v_{n}^{(\kappa)}$ and check the cancellation noted in \cite{peskin2}.

%\end{document}

%% file: BKp-surstate.tex
%\input definitions.tex
%\begin{document}

\dopage{\finkfile}

\subsection{Surface state $|0\rangle\star|0\rangle$}

One can consider a so called surface state
\cite{peskin2,universality,0105168} that has a form
\begin{equation}
\langle f,-q-Q|\equiv\langle-q-Q|U_{f} \, ,
\quad U_{f}=\exp\left(\sum_{n=1}^{\infty}v_{n}L_{n}\right)
\label{<f|}
\end{equation}
where $v_{n}$ are specify a vector field $v(z)=\sum_{n}v_{n}z^{n}$ specifying
a finite conformal transformation
$$
f(z)=e^{v(z)\partial_{z}}z
$$
(we also assumed a particular choice of the $SL(2,{\mathbb R})$-frame for which $f(0)=0$).

The $\bpz$-conjugated state is
$$
|I\circ f\circ I, -q-Q\rangle \equiv U^{-1}_{I\circ f \circ I}|-q-Q\rangle \,,\qquad
U^{-1}_{I\circ f \circ I}=\exp\left(\sum_{n=1}^{\infty}(-1)^{n}v_{n}L_{-n}\right)
$$
where $I$ stands for the inversion mapping: $I(z)=-1/z$.
The claim is that the scalar product of any two surface states in the
CFT with zero central charge is equal to $1$
\begin{equation}
\langle f, -Q|I\circ g\circ I,0\rangle  =1 \, .
\label{<fg>}
\end{equation}
There is a well known perturbative reason for this \cite{peskin2}.
If we expand in series the operator $U$ defined by \eqref{<f|} then
the scalar product \eqref{<fg>} will be given by a sum of the correlation
functions involving only the stress energy tensor.
In a theory with vanishing central charge all such correlators vanish,
and therefore we have only one non-zero term in the sum, which is equal to $1$.

Let us discuss in detail how we can check (non-perturbatively)
 \eqref{<fg>}  for the case of a special
surface state $|0\rangle\star |0\rangle$, which
corresponds to the map \cite{universality}
$$
f(z)=f^{(3)}_{1}\equiv \left(\frac{1-iz}{1+iz}\right)^{2/3} \,.
$$
Written in the oscillator representation  the ghost parts of the corresponding
surface
states read
\begin{subequations}
\begin{equation}
\langle f^{(3)}_1,-q-Q|=\langle -q-Q|\exp\left[
-\frac{\epsilon}{2}\sum_{n,m=1}^{\infty}\af_n V_{nm}^{\prime 11}\af_m
-\frac{\epsilon q}{\sqrt{2}}\sum_{n=1}^{\infty}V_{0n}^{\prime 11}\af_n
+\frac{\epsilon Q}{3\sqrt{2}}\sum_{n=1}^{\infty}J_n\af_n
\right] \label{ss1}
\end{equation}
where for generality we wrote the state for an arbitrary vacuum ghost
number \cite{peskin2}. Strictly speaking such a state with $q\ne 0$ should not be called a
surface state as it is not built over the conformal vacuum.
However insertions proportional to the background charge are always
needed to get a non-vanishing overlap between two surface states. The terms proportional
to $Q$ that occur in (\ref{ss1}), (\ref{ss2})   serve precisely that purpose.
(So we might as well call these states ``modified surface states'', or
``surface states with insertions''.)

The state $\bpz$-conjugated to (\ref{ss1})  is of the form
\begin{equation}
|I\circ f^{(3)}_1\circ I,-q-Q\rangle=\exp\left[
-\frac{\epsilon}{2}\sum_{n,m=1}^{\infty}\af_{n}^{\dag}V_{nm}^{\prime 11}\af_{m}^{\dag}
+\frac{\epsilon q}{\sqrt{2}}\sum_{n=1}^{\infty}V_{0n}^{\prime 11}\af_{n}^{\dag}
-\frac{\epsilon Q}{3\sqrt{2}}\sum_{n=1}^{\infty}J_n\af_n^{\dag}
\right]| -q-Q\rangle\, . \label{ss2}
\end{equation}
\end{subequations}

Now using the functional integral technique \cite{Berezin} we find
that the scalar product \eqref{<fg>} is
\begin{multline}
\langle f^{(3)}_{1}, -Q|I\circ f^{(3)}_{1}\circ I,0\rangle=
\\
=\det\left[1-(V^{\prime 11})^2\right]^{-\frac{D+1}{2}}
\exp\left[
\frac{\epsilon Q^2}{4}\,\left(
-\frac{2}{9}J_n\left([1-V^{\prime 11}]^{-1}
\right)_{nm}J_{m}
\right.\right.
\\
\left.\left.
-\frac{2}{3}V_{0n}^{\prime 11}\left([1-V^{\prime 11}]^{-1}
\right)_{nm}J_{m}
-V_{0n}^{\prime 11}\left([1-(V^{\prime 11})^2]^{-1}V^{\prime 11}
\right)_{nm}V_{0m}^{\prime 11}
\right)
\right] \, .
\end{multline}
One can further simplify the above expressions  using the identities
\begin{equation*}
\frac{1}{1-V^{\prime 11}}=\frac{1+3V^{\prime 11}}{4(1-V^{\prime 11})}+\frac{3}{4}
\quad\text{and}\quad V_{0n}^{\prime 11}=
-\frac{1}{3}\left(1+3V^{\prime 11}\right)_{nm}J_m
\end{equation*}
(see Appendix A).
Finally we get
\begin{multline}
\langle f^{(3)}_{1}, -Q|I\circ f^{(3)}_{1}\circ I,0\rangle
=\det\left[1-(V^{\prime 11})^2\right]^{-\frac{D+1}{2}}
\exp\left[
\frac{\epsilon Q^2}{4}\,\left(
-\frac{1}{6}\sum_{n=1}^{\infty}J_n^2\right.\right.
\\
\left.\left.
-\frac{1}{2}V_{0n}^{\prime 11}\left([1-V^{\prime 11}]^{-1}
\right)_{nm}J_{m}
%\right.\right.
%\left.\left.
-\sum_{n,m}V_{0n}^{\prime 11}\left([1-(V^{\prime 11})^2]^{-1}V^{\prime 11}
\right)_{nm}V_{0m}^{\prime 11}
\right)
\right]
\label{identity}
\end{multline}
Here we rearranged the terms in the exponent so that
the divergent part $\sum_{n=1}^{\infty}J_n^2$ is singled out. The
rest of the terms in the exponent are finite. This will be clear once we
will rewrite these terms in the $v_{n}^{(\kappa)}$ basis in Section~\ref{subsec:53}.
In that basis these terms are represented by convergent integrals.

Our goal now is to check that expression \eqref{identity} is equal to $1$
in the case of critical bosonic string, i.e. $\epsilon=1$ $Q=-3$ and $D=26$.
We would like to note that this identity is a particular case of  formula (5.60)
in \cite{peskin2}.

\subsection{Regularization  of  determinants. Cancellation of infinities.}
Consider a symmetric
operator $G_{nm}$ acting in space of infinite sequences $l_{\infty}$ labelled by the
string mode index $n=1,2,\dots$. Assume further that this operator takes a diagonal form
in the basis $v_{n}^{(\kappa)}$ and its eigenvalues are $G(\kappa)$.
Then its determinant can be formally written as
\begin{subequations}
\label{detab}
\begin{equation}
\det G_{nm}
=\exp\bigl(
\Tr\log G_{nm}\bigr)
=\exp\left[
\int_{-\infty}^{\infty}d\kappa\,
\rho(\kappa)\log G(\kappa) \right],
\label{deta}
\end{equation}
where $\rho(\kappa)$ is a spectral density.
If the operator $G_{nm}$ is rewritten in the basis of even and odd
eigenvectors, its determinant will have the following form\footnote{
Notice that if $v^{(\kappa)}$ are eigenvectors of a symmetric
operator $G_{nm}$
then $G_{\alpha\beta}(\kappa)=G(\kappa)\delta_{\alpha\beta}$ and
$G(-\kappa)=G(\kappa)$.}
\begin{equation}
\det G_{\alpha\beta}(\kappa)
=\exp\bigl(
\Tr\log G_{\alpha\beta}(\kappa)\bigr)
=\exp\left[
\int_{0}^{\infty}d\kappa\,
\rho(\kappa)\log G(\kappa) \right]
\label{detb}
\end{equation}
\end{subequations}
where $\rho(\kappa)$ is the same spectral density
as in \eqref{deta} and the indices $\alpha$, $\beta$ take  values ``o'', ``e''.

 It was suggested in \cite{spectroscopy},
 \cite{okuyama1} that the regularized spectral density
 is $\rho_L(\kappa)=\frac{\log L}{2\pi}$ where $L$ is a ``level regulator'' that
 truncates only the mode labels $m,n\leqslant L$.
Strictly speaking this defines only the divergent part of the spectral
density. The whole spectral density regulated by $L$ can be written as
\begin{equation}
\rho_{2L}(\kappa)=\frac{1}{\Nc(\kappa)}\sum_{n=1}^{2L}v_n^{(\kappa)}v_n^{(\kappa)}
=\frac{1}{2\pi}\sum_{n=1}^{L}\frac{1}{n}+\rho_{\text{fin}}^{2L}(\kappa)\, .\
\label{density}
\end{equation}
The first term in this expression  gives
(up to a finite constant)
the spectral density $\frac{\log L}{2\pi}$ used in \cite{spectroscopy},\cite{okuyama1}.
The second term is  finite in the limit $L\to \infty$. We will discuss it in more detail
in the next subsection concentrating for now on the divergent part.

Using \eqref{detab}, (\ref{density})
we can write the following expression for the determinant
appearing in \eqref{identity}
\begin{multline}
\det\left[1-(V^{\prime 11})^2\right]
=\exp\left[
\int_{-\infty}^{\infty}d\kappa\,
\rho(\kappa)\log\bigl(1-\mu^2\bigr)\right]
\\
=\lim_{L\to\infty}\exp\left[-\frac{1}{36}\sum_{n=1}^{L}\frac{1}{n}
+\int_{-\infty}^{\infty}d\kappa\,
\rho_{\text{fin}}^{2L}(\kappa)\log\bigl(1-\mu^2\bigr)\right].
\label{det}
\end{multline}
To obtain this equation we have used
\begin{equation*}
\int_{-\infty}^{\infty}d\kappa\,
\log\bigl(1-\mu^2\bigr)=-\frac{\pi}{18}.
\end{equation*}

The divergent part in the exponential term standing in (\ref{identity}) is
\begin{equation}
\exp\Bigl[ -\frac{\epsilon Q^{2}}{24}\sum_{n=1}^{\infty}J_{n}^{2}\Bigr] \sim
\exp\Bigl[ -\frac{\epsilon Q^{2}}{24}\sum_{n=1}^{L}\frac{1}{n}\Bigr] \, .
\label{expp}
\end{equation}
Collecting both divergent terms from (\ref{det}), (\ref{expp}) we obtain for
the divergent part  of the expression (\ref{identity})
$$
\exp\Bigl[ \left(\frac{D+1}{72}-\frac{\epsilon Q^2}{24}\right)\sum_{n=1}^{L}\frac{1}{n} \Bigr]
$$
that indeed vanishes when we plug in the parameters corresponding to the critical bosonic
string theory ($D=26$, $\epsilon=1$ and $Q=-3$).

\subsection{ Finite parts}
\label{subsec:53}

\begin{figure}[t]
\centering
\includegraphics[width=300pt]{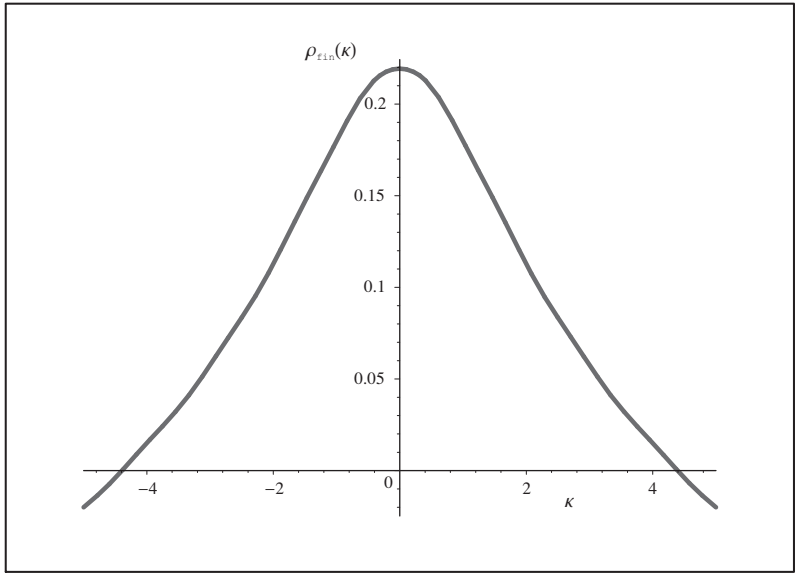}
\caption{Plot of the finite part of the spectral density $\rho_{\text{fin}}^{2L}(\kappa)$
 computed at the level $L=62$. $\rho_{\text{fin}}(0)=\frac{\log 2}{\pi}$.}
\label{fig:rho}
\end{figure}

To compute the finite part of the determinant (\ref{det}) we need to know the
finite part $\rho_{\text{fin}}(\kappa)$ of the spectral density.
So far we could not find a closed analytic expression for $\rho_{\text{fin}}(\kappa)$ and
will present here only some numeric results. Figure~\ref{fig:rho} gives a plot
of $\rho_{\text{fin}}^{2L}(\kappa)$ in the vicinity of $0$ computed for $L=62$.

The integral in the r.h.s. of \eqref{det}  converges rather quickly  and
can be easily computed numerically. We compute this integral
on the interval $(-30,30)$ for several values of $L$ (see Table~\ref{tab:1}).
All this values are in a good agrement this the
interpolating formula $a+b/\log L$.
\begin{table}[!h]
\centering
\renewcommand{\arraystretch}{1.5}
\begin{tabular}{||C|C|C|C|C|C||}
\hline
L & 32 & 52 & 62 &  82 & \infty
\\
\hline
\int  & -0.03550 &  -0.03558
& -0.03561 &  -0.03563 & -0.03602
\\
\hline
\end{tabular}
\caption{Here we present the numeric values
of the quickly convergent integral
$\int_{-\infty}^{\infty}d\kappa\,\rho_{\text{fin}}^{2L}(\kappa)
\log\bigl(1-\mu^2\bigr)$ computed at level $L$. In the very right column
we present the value for this integral estimated by the
formula $a+b/\log L$.}
\label{tab:1}
\end{table}

The finite part of the exponent in \eqref{identity}  can be computed by
summing up residues in the upper half
plane
\begin{multline}
(\text{finite})=\frac{\epsilon Q^2}{4}\,\left(
-\frac{1}{2}V_{0n}^{\prime 11}\left([1-V^{\prime 11}]^{-1}
\right)_{nm}J_{m}
-\sum_{n,m}V_{0n}^{\prime 11}\left([1-(V^{\prime 11})^2]^{-1}V^{\prime 11}
\right)_{nm}V_{0m}^{\prime 11}
\right)
\\
=\frac{\epsilon Q^2}{12}\int_{-\infty}^{\infty}\frac{d\kappa}{\Nc(\kappa)}\frac{2}{\kappa^2}
\frac{1+3\mu}{1-\mu}\left[\frac{1}{2}-\frac{1}{3}\frac{1+3\mu}{1+\mu}\mu\right]
\\
=-\frac{\epsilon Q^2}{3}\log\gamma-\frac{\epsilon Q^2}{36}
\left[\log 4+\frac{7}{2\pi^2}\zeta(3)\right].
\label{res2}
\end{multline}

If we  substitute all of our results \eqref{det}, \eqref{res2} into
equation \eqref{identity} we obtain
\begin{multline}
\log\,\langle f^{(3)}_{1}, -Q|I\circ f^{(3)}_{1}\circ I,0\rangle=
\lim_{L\to\infty}
\left(\frac{D+1}{72}-\frac{\epsilon Q^2}{24}\right)\sum_{n=1}^{L}\frac{1}{n}
\\
-\frac{D+1}{2}\int_{-\infty}^{\infty}d\kappa\,\rho_{\text{fin}}^{2L}(\kappa)
\log(1-\mu^2)
-\frac{\epsilon Q^2}{3}\log\gamma-\frac{\epsilon Q^2}{36}
\left[\log 4+\frac{7}{2\pi^2}\zeta(3)\right]
\end{multline}
where for completeness we wrote both divergent and finite parts. As already
noted the divergent parts cancel out for $Q=-3$, $D=26$, $\epsilon=1$.
The value of the finite part can be estimated using the results presented
in Table~\ref{tab:1}. It turns out to be close to the number $0.81805$
that seams to be far  from the wanted $0$. We hope that the origin
of this difference is just an  artifact of our numeric calculations.
 An analytic expression for  the finite part
of the spectral density  would  certainly come handy in proving  the exact identity
\eqref{<fg>}.

%\end{document}

%% file: BKp-iso.tex
%\input definitions.tex
%\begin{document}

\dopage{\finkfile}

\subsection{Normalization of the multiplication kernel}
The total multiplication kernel combining the matter and ghost sectors
contains an overall normalization constant
\begin{equation}
\Ks_{Q=0}^{D}\Ks_Q=\det\Bigl[\frac{2}{\theta^{2}(\kappa)}  \Bigr]^{D+1}
\times \det\Bigl[\frac{12+\theta^{2}(\kappa)}{16}  \Bigr]^{D+1}
\exp\Bigl[\epsilon\frac{Q^{2}}{24}\sum_{n=1}^{\infty}J_{n}^{2} -
\epsilon\frac{Q^{2}}{3}\log\gamma   \Bigr] \, .
\end{equation}

 We can now apply the regularization technique of the previous section to
compute this constant. We will keep the first factor untouched for the reason to
be discussed a little later. The divergent  part contained in  the second two factors
reads
\begin{equation}
\exp\left[
\left(\frac{D+1}{2\pi}\int_{0}^{\infty}d\kappa\,
\log\left(\frac{12+\theta^2(\kappa)}{16}\right)
+\frac{\epsilon Q^2}{24}\right)
\sum_{n=1}^{L}\frac{1}{n} \right] =
\exp\left[
\left(-\frac{D+1}{18}
+\frac{\epsilon Q^2}{24}\right)
\sum_{n=1}^{L}\frac{1}{n} \right]
\end{equation}
and we see that for the parameters corresponding to the critical
bosonic string these infinities do not cancel each other.
This agrees with an analogous computation made in \cite{Erler}.
In this calculations we use
\begin{equation*}
\int_{0}^{\infty}d\kappa\,
\log\left(\frac{12+\theta^2(\kappa)}{16}\right)=-\frac{\pi}{9}.
\end{equation*}

We must note here that there is a potential subtlety in the above computation having to do
with the finite part of the spectral density. It contributes the exponent
$$
\exp\Bigl[(D+1)\int_{0}^{\infty}d\kappa\,\rho_{\text{fin}}(\kappa)
\log\left(\frac{16}{12+\theta^2(\kappa)}\right)     \Bigr]
$$
which we attributed to the finite part. Strictly speaking we do not know
the asymptotics of the function $\rho_{\text{fin}}(\kappa) $ as $\kappa$ goes
to infinity. However given the fact that it is being integrated  with a factor that is
exponentially falling off at infinity it looks  unlikely that the integral diverges.
We hope to clarify this subtlety in a future work.

%%%%%%%%%%%%%%%%%%
%$\Cc$ is a constant, which we are going to determine
%and $\theta(\kappa)$ is given by \eqref{theta}.
%Using the equations \eqref{iso}, \eqref{kernelQ}, \eqref{M} and
%\eqref{kernelM} one can write the following expression for
%the constant~$\Cc$
%\begin{equation}
%\Cc=\frac{\det\bigl[\pi\theta(\kappa)\bigr]^{-2(D+1)}}{\Ks_{Q=0}^{D}\Ks_Q}
%=\det\left[\frac{1}{2\sqrt{\pi}}\,\frac{16}{12+\theta^2(\kappa)}\right]^{D+1}
%\exp\left[-\frac{\epsilon Q^2}{24}\sum_{n=1}^{\infty}J_n^2
%+\frac{\epsilon Q^2}{3}\log\gamma\right]
%\end{equation}

%%%%%%%%%%%%%%%%%%%%%%%%%%%%%%%%%%

\subsection{Relation between continuous Moyal and string products}

In this section we will try to state clearly the precise correspondence between Witten's and
continuous Moyal star products.
 We start by reminding that the Moyal product of two functions
on $\Rh^2$ can be defined by a kernel $K_{\theta}(x_1,x_2,x_3)$
in the following way
\begin{subequations}
\begin{equation}
(f\star g)(x_1)=\int_{\Rh^2}dx_2dx_3\,
K_{\theta}(x_1,x_2,x_3)f(x_2)g(x_3),
\label{mfunc}
\end{equation}
where
\begin{equation}
K_{\theta}(x_1,x_2,x_3)=\frac{1}{\pi^2\theta^2}\,\exp\left[
-\frac{2i}{\theta}\epsilon_{\alpha\beta}\Bigl(
x_1^{\alpha}x_2^{\beta}+
x_2^{\alpha}x_3^{\beta}+
x_3^{\alpha}x_1^{\beta}
\Bigr)
\right]
\label{moyal}
\end{equation}
\end{subequations}
$\theta$ is a real deformation parameter and
matrix $\epsilon_{\alpha\beta}$ is defined in \eqref{matrices}. Notice that
one needs the factor $\theta^{-2}$  in the kernel
to obtain a kernel for  pointwise  multiplication in the limit $\theta\to 0$.

The continuous Moyal product will be defined
as a product of functionals $\Fs[X^{\alpha,\mu}(\kappa)]$, where the functions
$X^{\alpha,\mu}(\kappa):\Rh_+
\to \Rh^2\otimes \Rh^D$
($\alpha=e,o$ and $\mu=0,\dots,D-1$) will be considered as canonical  coordinates
on our non-commutative space. Then the product can be defined as follows
\begin{subequations}
\begin{equation}
(\Fs_1\ast\Fs_2)[X^{(3)}]=\int\Ds X^{(1)}\Ds X^{(2)}\,
K_{\text{Moyal}}\bigl(X^{(1)},X^{(2)},X^{(3)}\bigr)\Fs_1[X^{(1)}]\Fs_2[X^{(2)}],
\label{Mfunc}
\end{equation}
where the measure $\Ds X$ is defined in \eqref{measure}
and the continuous Moyal kernel $K_{\text{Moyal}}$ is
of the form
\begin{equation}
K_{\text{Moyal}}\bigl(X^{(1)},X^{(2)},X^{(3)})=
\det\bigl[\theta(\kappa)\bigr]^{-2D}
\exp\left[2i \int_{0}^{\infty}\frac{d\kappa}{\theta(\kappa)}\,
\sum_{\mu=0}^{D-1}x_{e,\mu}^{(r)}(\kappa)\chi^{rs}x_{o,\mu}^{(s)}(\kappa)\right].
\label{kernelM}
\end{equation}
\end{subequations}
Here the determinant should  be understood as in \eqref{detb}.
We also include in this definition a normalization factor $\det\bigl[\theta(\kappa)\bigr]^{-2D}$.
Despite the fact that it is an infinite quantity this factor is needed
to obtain a correct limit as $\kappa \to 0$, when we get a commutative mode.
One might think that the kernels \eqref{kernelM} and \eqref{moyal}
differ due to the factor $\pi^2$, but actually this difference
is only because of the difference in the measure normalization
we use in \eqref{Mfunc} and \eqref{mfunc}.

The combined multiplication kernel that we obtained differs from (\ref{kernelM})
by a linear exponent factor present in the ghost kernel (\ref{kernel_q}) and by an
additional normalization factor $\Cc$:
\begin{equation}
\Cc=\frac{\Ks_{Q=0}^{D}\Ks_Q}{\det\bigl[\theta(\kappa)\bigr]^{-2(D+1)}}
=\det\left[2\cdot\frac{12+\theta^2(\kappa)}{16}\right]^{D+1}
\exp\left[-\frac{\epsilon Q^2}{24}\sum_{n=1}^{\infty}J_n^2
+\frac{\epsilon Q^2}{3}\log\gamma\right]
\label{constant}
\end{equation}

Our results can therefore by summarized in the following mapping establishing
an isomorphism of algebras
\begin{equation}
{\cal I}: \Cc^{-1}\cdot \exp\left[-i\sqrt{2}\epsilon\Bigl(q+\frac{Q}{2}
\Bigr)\int_{0}^{\infty}d\kappa\,
 J_{\kappa} x_{e}(\kappa)\right]\Psi_{q}[X]\quad\mapsto \quad
 \Fs_q[X],
\label{iso}
\end{equation}
that is ${\cal I}$ specifies  a field redefinition that maps the Witten star
product into the canonically normalized continuous Moyal product (\ref{kernelM}).
In view of the investigations made in the previous sections the redefinition involves
an infinite multiplicative factor (same problem was noted in the matter sector
in \cite{0202087}). We could in principle remove this factor from the kernel and hide
it into the $\Ds X$ functional integration measure, but then it would show
up again in front of  normalized wave functionals, like the one
representing the vacuum state. We are not sure though at the current stage of investigation
whether this infinite factor is a serious drawback of the continuous Moyal formalism in
SFT. One should be able to do computations  keeping these factors   regulated and finite.

%\end{document}

%% file: BKp-appA.tex
%\input definitions.tex
%\begin{document}

\dopage{\finkfile}

The generating function for the vector $J_n$ can be
obtained directly from the expression \eqref{Jn} and
is of the form
\begin{equation}
J(z)=\sum_{n=1}^{\infty}J_n\frac{z^n}{\sqrt{n}}
=-\frac{1}{\sqrt{2}}\log(1+z^2).
\end{equation}
Now comparing this function with the generating function (B.26) from \cite{dima1}
and using the expressions (3.17a) and (B.21) from \cite{dima1}
one can obtain the following representation for the vector $J$
\begin{equation}
J=-\mathscr{P}\frac{1}{C'U'+1}W.
\label{JW}
\end{equation}
Notice also that vector $J$ is an even one
\begin{equation}
C'J=J.
\end{equation}

Using the representation \eqref{JW} one can obtain the following
useful relations involving vector $J$
\begin{subequations}
\begin{align}
U^{\prime}J&=-\bar{W}-J
\\
\bar{U}^{\prime}J&=-W-J;
\end{align}
\label{J-prop}
\end{subequations}

We also need to know the inner product of $J$ with $W$
\begin{equation}
\sum_{n=1}^{\infty}W_nJ_n=\sum_{n=1}^{\infty}\bar{W}_nJ_n=-V_{00}'.
\end{equation}
This relation can be obtained using diagonal representation
of $J$ and $W$
\begin{multline}
\sum_{n=1}^{\infty}W_nJ_n=-\int_{-\infty}^{\infty}\frac{d\kappa}{\Nc(\kappa)}
\,\frac{2}{\kappa^2}(1+\nu(\kappa))
\equiv
-\int_{-\infty}^{\infty}\frac{d\kappa}{\kappa^2\Nc(\kappa)}
\,\bigl[2+\nu(\kappa)+\bar{\nu}(\kappa)\bigr]
\\
=-2\int_{0}^{\infty}\frac{d\kappa}{\kappa^2\Nc(\kappa)}
\,\bigl[1+3\mu(\kappa)\bigr]=-V_{00}'
\label{V_00}
\end{multline}

%\end{document}

%% file: BKp-appB.tex
%\input definitions.tex
%\begin{document}

\dopage{\finkfile}

Let us rewrite the terms appearing in the exponential
\eqref{V3} in the basis $\ef_{\kappa}^{\dag}$, $\of_{\kappa}^{\dag}$.
%\begin{itemize}
%\item
 The quadratic part gets the following from in the diagonal basis
\begin{subequations}
\begin{equation}
\sum_{r,s}\sum_{n,m=1}^{\infty}a_{n}^{(r)\dag}V_{nm}^{\prime\,rs}a_{m}^{(s)\dag}=
\int_{-\infty}^{\infty}d\kappa\,C'a_{\kappa}^{(r)\dag}\mu^{rs}(\kappa)a_{\kappa}^{(s)\dag}
=\int_{0}^{\infty}d\kappa\,
a_{\kappa,\alpha}^{(r)\dag}
V^{rs}{}_{\alpha\beta}
a_{\kappa,\beta}^{(s)\dag},
\end{equation}
where $a_{\kappa,\alpha}^{\dag}=(e_{\kappa}^{\dag},o_{\kappa}^{\dag})_{\alpha}$ and
\begin{align}
V^{rs}{}_{\alpha\beta}&=\mu \delta_{\alpha\beta}\otimes \delta^{rs}
+\mu_s \delta_{\alpha\beta}\otimes\varepsilon^{rs}
+i\mu_a\epsilon_{\alpha\beta}\otimes\chi^{rs};
\\
\mu&=\mu^{11}=\frac{-1+t^2}{3+t^2},\qquad
\mu_s=\frac12(\mu^{12}+\mu^{21})=\frac{2}{3+t^2}
\\
\mu_a&=\frac12(\mu^{12}-\mu^{21})=\frac{2t}{3+t^2},
\qquad t=\tanh\frac{\pi\kappa}{4}.
\end{align}
\end{subequations}
The matrices $\epsilon_{\alpha\beta}$, $\varepsilon^{rs}$ and $\chi^{rs}$
are defined in \eqref{matrices}.

%\item
The part linear in $a_n$ and $q$ gets the form
\begin{multline}
\sum_{r,s}\sum_{n=1}^{\infty}q^{(r)}V_{0n}^{\prime\,rs}a_n^{(s)\dag}
=-\frac{\sqrt{2}}{3}\int_{-\infty}^{\infty}
\frac{d\kappa}{\kappa\sqrt{\Nc(\kappa)}}
q^{(r)}\Bigl(3\delta^{rs}-2+3\mu^{rs}\Bigr)a_{\kappa}^{(s)\dag}
\\
=\int_{0}^{\infty}\frac{d\kappa}{\kappa\sqrt{\Nc(\kappa)}}
\left[
-(1+3\mu)\Bigl(q^{(r)}+\frac{Q}{3}\Bigr)e_{\kappa}^{(s)\dag}
-2i\mu_a q^{(r)}\chi^{rs}o_{\kappa}^{(s)\dag}
\right]
\\
=\int_{0}^{\infty}\frac{d\kappa}{\kappa\sqrt{\Nc(\kappa)}}
\, 2\mu_a
\left[
-t\Bigl(q^{(r)}+\frac{Q}{3}\Bigr)e_{\kappa}^{(s)\dag}
-iq^{(r)}\chi^{rs}o_{\kappa}^{(s)\dag}
\right]
\end{multline}

%\item
 The term corresponding to the midpoint insertion has the following
form in the diagonal basis
\begin{equation}
\sum_{n=1}^{\infty}J_n a_n^{(r)\dag}=
\sqrt{2}\int_{-\infty}^{\infty}\frac{d\kappa}{\kappa\sqrt{\Nc(\kappa)}}\,
a_{\kappa}^{(r)\dag}
\stackrel{\ref{ae}}{=}
2\int_{0}^{\infty}\frac{d\kappa}{\kappa\sqrt{\Nc(\kappa)}}\,e_{\kappa}^{(r)\dag}.
\end{equation}

%\item
The divergent term reads
\begin{equation}
\sum_{n=1}^{\infty}J_n^2=\int_{-\infty}^{\infty}\frac{d\kappa}{\Nc(\kappa)}\,
(v^{(\kappa)},J)^2=2\int_{-\infty}^{\infty}\frac{d\kappa}{\kappa^2\Nc(\kappa)}
=2\int_{0}^{\infty}d\kappa\,J_{\kappa}^2,
\end{equation}
where $J_{\kappa}$ is defined by \eqref{Jk}

%\item
The term that depends only on the momentum   can be rewritten in the integral
representation using expression \eqref{V_00}
\begin{equation}
V_{00}'\left(\sum_{r}q_r^2-Q^3\right)
=\int_0^{\infty}d\kappa\,J_{\kappa}^2(1+3\mu)\left(\sum_{r}q_r^2-Q^2\right).
\end{equation}
%\end{itemize}

%\end{document}

%% file: BKp-appC.tex
%\input definitions.tex
%\begin{document}

\dopage{\finkfile}

For simplicity we will calculate the kernel $K$
in basis \eqref{basisX} for fixed $\kappa$. This means
that we will drop integration over $\kappa$ in the proceeding calculations.
The kernel defining multiplication in the coordinate representation
has the following form (we assume $\sum_r q_r+Q=0$)
\begin{multline}
K_{\{q_r\}}(X^{(1)},X^{(2)},X^{(3)})=\bigl(
\langle X^{(1)}_{\kappa}|\otimes\langle X^{(2)}_{\kappa}|\otimes\langle X^{(3)}_{\kappa}|
\bigr)|V_3\rangle
\\
=\int \mathscr{D}a\mathscr{D}a^*\,
\exp\left[
-\frac{\epsilon}{2}
\begin{pmatrix}
a & a^*
\end{pmatrix}
\begin{pmatrix}
-1 & 1\\
1& V
\end{pmatrix}
\begin{pmatrix}
a \\
a^*
\end{pmatrix}
-\epsilon \begin{pmatrix}
-i\sqrt{2}x & \Phi
\end{pmatrix}
\begin{pmatrix}
a \\
a^*
\end{pmatrix}
\right]
\\
\times\exp\left[
-\frac{\epsilon}{2}\sum_r (x^{(r)},x^{(r)})
-\frac{\epsilon}{4}J_\kappa^2(1+3\mu)\Bigl(
\sum_r q_r^2-Q^2
\Bigr)
\right]
\\
=\det{}^{-1/2}(1+V)\exp\left[
\frac{\epsilon}{2}
\begin{pmatrix}
-i\sqrt{2}x & \Phi
\end{pmatrix}
\begin{pmatrix}
-1 & 1\\
1& V
\end{pmatrix}^{-1}
\begin{pmatrix}
-i\sqrt{2}x\\
 \Phi
\end{pmatrix}
\right]
\\
\times\exp\left[
-\frac{\epsilon}{2}\sum_r (x^{(r)},x^{(r)})
-\frac{\epsilon}{4}J_\kappa^2(1+3\mu)\Bigl(
\sum_r q_r^2-Q^2
\Bigr)
\right]
\end{multline}
Notice that
\begin{equation*}
\begin{pmatrix}
-1 & 1\\
1& V
\end{pmatrix}^{-1}
=(1+V)^{-1}\begin{pmatrix}
1 & 1\\
1& 1
\end{pmatrix}
+
\begin{pmatrix}
-1 & 0\\
0& 0
\end{pmatrix}
\end{equation*}
and
\begin{equation}
(1+V)^{-1}=\frac{1}{2}\otimes 1-\frac{i}{2t}\epsilon\otimes \chi.
\end{equation}
Substitution yields the following expression for the kernel
\begin{multline}
K_{\{q_r\}}(x^{(1)},x^{(2)},x^{(3)})=\det{}^{-1/2}(1+V)\exp
\left[
\frac{\epsilon}{2}\sum_r(x^{(r)},x^{(r)})-\epsilon\, x(1+V)^{-1}x
\right.
\\
\left.
-i\epsilon\sqrt{2}\,x(1+V)^{-1}\Phi+\frac{\epsilon}{2}\Phi(1+V)^{-1}\Phi\
-\frac{\epsilon}{4}J_{\kappa}^2(1+3\mu)\Bigl(\sum_r q_r^2-Q^2\Bigr)
\right]
\label{Kp}
\end{multline}
We further obtain
\begin{subequations}
\label{xxPP}
%\begin{itemize}
%\item
the $xx$-term:
\begin{equation}
x(1+V)^{-1}x=\frac{1}{2}\sum_r(x^{(r)},x^{(r)})-\frac{i}{t}\Bigl(
x^{(1)}\wedge x^{(2)}+x^{(2)}\wedge x^{(3)}+x^{(3)}\wedge x^{(1)}
\Bigr) \, ,
\end{equation}
%\item
the  $x\Phi$-term:
\begin{multline}
\sqrt{2}\,x(1+V)^{-1}\Phi=\frac{1}{\sqrt{2}}\sum_r\left(
x_e^{(r)}\Phi_e^{(r)}+x_o^{(r)}\Phi_o^{(r)}
\right)
-\frac{1}{\sqrt{2}}\frac{i}{t}\left(
\Phi_e^{(r)}\chi^{rs}x_o^{(s)}-\Phi_o^{(r)}\chi^{rs}x_e^{(s)}
\right)
\\
\stackrel{\ref{Phioe}}{=}
\frac{1}{\sqrt{2}}\left[
x_e^{(r)}\Phi_e^{(r)}+\frac{i}{t}\Phi_o^{(r)}\chi^{rs}x_e^{(s)}
\right]
=\sqrt{2}J_{\kappa}\sum_r x_e^{(r)}\Bigl(q^{(r)}+\frac{Q}{2}\Bigr)\, ,
\end{multline}
%\item
the $\Phi\Phi$-term:
\begin{multline}
\Phi(1+V)^{-1}\Phi=\frac{1}{2}\Bigl(
\Phi_e^{(r)}\Phi_e^{(r)}+\Phi_o^{(r)}\Phi_o^{(r)}
\Bigr)
-\frac{i}{t}\Phi_e^{(r)}\chi^{rs}\Phi_o^{(s)}
\\
\stackrel{\ref{Phioe}}{=}\frac{1}{2}\Bigl(
\Phi_e^{(r)}\Phi_e^{(r)}-\Phi_o^{(r)}\Phi_o^{(r)}
\Bigr)
=\frac{1}{2}J_{\kappa}^2(1+3\mu)\left[
\sum_r q_r^2-\frac{Q^2}{3}
\right]+\frac{Q^2}{6}J_{\kappa}^2
\end{multline}
%\item
while the determinant reads
\begin{equation}
\label{Deta}
\det(1+V)^{-1}=\det(\frac{1}{2}\otimes 1-\frac{i}{2t}\epsilon\otimes\chi)
=\frac{1}{2^6}\det\left[1\otimes 1-\frac{i}{t}\epsilon\otimes\chi\right]\, .
\end{equation}
We used the following trick to calculate the determinant:
\begin{multline}
\label{Detb}
\log \det\left[1\otimes 1-\frac{i}{t}\epsilon\otimes\chi\right]=
\Tr\log\left[1\otimes 1-\frac{i}{t}\epsilon\otimes\chi\right]
=-\Tr\sum_{n=1}^{\infty}\frac{(-1)^n}{n}\frac{(-i)^n}{t^n}\epsilon^n\otimes \chi^n
\\
=-\sum_{n=1}^{\infty}\frac{1}{n}\,t^{-2n}\Tr\chi^{2n}
=-\sum_{n=1}^{\infty}\frac{1}{n}\,t^{-2n}(-3)^{n-1}(-2\cdot 3)
=2\log\left[1+\frac{3}{t^2}\right]
\end{multline}
To obtain this we use $\chi^{2n}=(-3)^{n-1}(\varepsilon-2)$,
$\epsilon^{2n}=(-1)^n \mathbf{1}_2$. Combining equations \eqref{Deta}
and \eqref{Detb}
one obtains the following formula for the determinant
\begin{equation}
\det(1+V)=4\left(\frac{4t^2}{t^2+3}\right)^2\equiv 4(1+3\mu)^2
\end{equation}
%\end{itemize}
\end{subequations}
Substitution of \eqref{xxPP} into \eqref{Kp} yields
\begin{multline*}
K_{\{q_r\}}(x^{(r)})=\frac{1}{2(1+3\mu)}\exp\left[
\frac{i\epsilon}{t}\Bigl(x^{(1)}\wedge x^{(2)}+x^{(2)}\wedge x^{(3)}+x^{(3)}\wedge x^{(1)}\Bigr)
\right.
\\
\left.
-2i\epsilon\lambda_{\kappa}\sum_r x_e^{(r)}\Bigl(
q^{(r)}+\frac{Q}{2}
\Bigr)
+\left(\frac{\epsilon Q^2}{24}\sum_{n=1}^{\infty}J_n^2-\frac{\epsilon Q^2}{3}\log\gamma
\right)
\right]
\end{multline*}

%\end{document}

%% file: BKp-appD.tex
%\input definitions.tex
%\begin{document}

\dopage{\finkfile}
The aim of this appendix is to adapt the formulae
obtained in Sections~\ref{sec:diag} and \ref{sec:mult}
to the non-zero momentum $3$-string matter vertex.
Essentially all we need to do is to put $Q=0$,
substitute $p_r=-q_r$, change the Kronecker symbol
to the Dirac delta function, change all sums to the integrals
and substitute $\epsilon\mapsto g_{\mu\nu}$.

The matter $3$-string vertex in the diagonal basis
has the following form
\begin{multline}
|V_{3}^{(\text{matter})}\rangle_{123}=\frac{1}{(2\pi)^{\frac{D}{2}}}
\int d^Dp^{(1)}d^Dp^{(2)}d^Dp^{(3)}\,(\deg g)^{-1}
\delta(p^{(1)}+p^{(2)}+p^{(3)})
\\
\times
\exp\left[\int_{0}^{\infty}d\kappa\left(
-\frac{1}{4}J_{\kappa}^2(1+3\mu)
p^{(r)}_{\mu}g^{\mu\nu}p^{(r)}_{\nu}
\Bigr]\right. \right. \\
\left.\left.
-\sum_r\Phi^{(r)}_{\alpha,\mu}(\kappa)g^{\mu\nu}\af_{\kappa,\alpha,\mu}^{(r)\dag}
-\frac{1}{2}g^{\mu\nu}\af_{\kappa,\alpha,\mu}^{(r)\dag}V^{rs}_{\kappa,\alpha\beta}
\af_{\kappa,\beta,\nu}^{(s)\dag}
\right)
\right] \bigotimes_{r=1}^{3}|p^{(r)}\rangle,
\end{multline}
where $V_{\kappa,\alpha\beta}^{rs}$ is the same as for the ghost part and
\begin{align}
\Phi_{e,\mu}^{(r)}(\kappa)&=-\frac{1}{2}J_{\kappa}(1+3\mu)
p^{(r)}_{\mu},
\\
\Phi_{o,\mu}^{(r)}(\kappa)&=-i J_{\kappa}\mu_a p^{(s)}_{\mu}\chi^{sr}
\equiv\frac{2i}{\theta(\kappa)}\,\Phi_e^{(s)}\chi^{sr}.
\end{align}

The multiplication kernel in the mixed coordinate/momentum
basis has the form
\begin{multline}
K_{\{p_1,p_2\}}(X^{(r)})\equiv\Bigl(\langle X^{(1)},p_1|\otimes
\langle X^{(2)},p_2|\otimes\langle X^{(3)},-p_1-p_2|\Bigr)|V_3\rangle
\\
=
\Ks_{Q=0}^D
\exp\left[
2i\int_0^{\infty}\frac{d\kappa}{\theta(\kappa)}\,
g^{\mu\nu}x^{(r)}_{e,\mu}(\kappa)\chi^{rs} x^{(s)}_{o,\nu}(\kappa)
\right]\times
\\
\exp\left[i\sqrt{2}\int_0^{\infty}d\kappa\,
J_{\kappa}\Bigl\{ x_{e,\mu}^{(1)}(\kappa)
p_1^{\mu} + x_{e,\mu}^{(2)}(\kappa)p_2^{\mu}
- x_{e,\mu}^{(3)}(\kappa)\bar{p}_3^{\mu}\Bigr\}
\right]
\end{multline}
where $\bar{p}_3=p_1+p_2$.

%\end{document}